\documentclass[12pt]{article}
\usepackage{jhep-mod}
\usepackage{bm}
\usepackage{amssymb,amsmath,amsthm}
\usepackage{mathrsfs}
\usepackage{color}
\usepackage[utf8]{inputenc}
\usepackage{enumerate}
\usepackage{hyperref}
\hypersetup{colorlinks=true} 
\usepackage{appendix}
\usepackage{graphicx}
\usepackage{dcolumn}
\usepackage{bm}
\usepackage{multirow}
\usepackage{float}
\usepackage{xcolor}
\usepackage{tikz}
\definecolor{purple}{rgb}{1,0,1}
\definecolor{lime}{HTML}{A6CE39} 

\newcommand{\orcidicon}{%
	\begin{tikzpicture}
	\draw[lime, fill=lime] (0,0) 
		circle [radius=0.16] 
		node[white] {{\fontfamily{qag}\selectfont \tiny ID}};
	\draw[white, fill=white] (-0.0625,0.095) 
		circle [radius=0.007];
	\end{tikzpicture}
	\hspace{-3mm}
}
\newcommand\orcidDel{{\href{https://orcid.org/0000-0003-4158-202X}{\orcidicon}}}
\newcommand\orcidMatt{{\href{https://orcid.org/0000-0003-1088-6485}{\orcidicon}}}
\begin{document}
\title{\huge Explicit construction of the density matrix in Gleason's theorem}
\author{\Large Del Rajan\orcidDel{} {\sf and} Matt Visser\orcidMatt{}}
\emailAdd{del.rajan@sms.vuw.ac.nz, delrajan30@gmail.com}
\affiliation{School of Mathematics and Statistics, Victoria University of Wellington, \\
\null\quad PO Box 600, Wellington 6140, New Zealand}
\emailAdd{\break\null\quad\quad\quad\, matt.visser@sms.vuw.ac.nz}
\abstract{

\noindent
Gleason's theorem is a fundamental 60 year old result in the foundations of quantum mechanix, setting up and laying out the surprisingly minimal assumptions required to deduce the existence of quantum density matrices and the Born rule. Now Gleason's theorem and its proof have been continuously analyzed, simplified, and revised over the last 60 years, and we will have very little to say about the theorem and proof themselves. Instead, we find it useful, (and hopefully interesting), to make some clarifying comments concerning the explicit construction of the quantum density matrix that Gleason's theorem proves exists, but that Gleason's theorem otherwise says relatively little about. 

\noindent
\bigskip

\noindent
{\sc Pacs:} 03.65.-w; 03.65.Aa; 03.65.Ta

\bigskip
\noindent
{\sc Keywords:} \\
Gleason's theorem, probability functions, Hilbert subspaces, density matrix, Born rule, quantum probability, quantum ontology, quantum realism.

%
\bigskip
\noindent
{\sc Dated:} 1 April 2019; \LaTeX-ed \today

}

\maketitle
\theoremstyle{definition}
\newtheorem{lemma}{Lemma}
\parindent0pt
\parskip7pt
\def\C{\mathcal{C}}
\def\tr{{\mathrm{tr}}}
\def\d{{\mathrm{d}}}
\def\g{{\mathfrak{g}}}
\def\dbar{{\mathchar'26\mkern-12mu \d}} 
\def\Hilbert{{\mathcal{H}}}
\def\H{{\mathrm{H}}}
\def\R{{\mathrm{R}}}
\def\E{{\mathrm{E}}}
\def\nn{\nonumber}
\section{Introduction}
Gleason's theorem has a long 60-year-old history~\cite{Gleason,Cooke,Hellman,Billinge,Pitowsky,Richman,Richman2,Busch,Caves,without,Wright}, and is regarded as one of the fundamental theorems in the foundations of quantum mechanix. The theorem addresses the minimal, (in  fact, quite  surprisingly minimal), assumptions required to deduce the existence of a quantum density matrix, (a unit trace Hermitian matrix encoding the notion of quantum probability), and Gleason's theorem pragmatically underlies the theoretical justification for adopting the Born rule in standard quantum mechanix.

With such a long history~\cite{Gleason,Cooke,Hellman,Billinge,Pitowsky,Richman,Richman2,Busch,Caves,without,Wright}, and in view of the publication of a number of recent related books~\cite{Isham, Heino, Chiara, Hamhalter,Cohen}, it is perhaps surprising that there is anything left to say on this subject.
Early proofs of Gleason's theorem were implicit and non-constructive, and for some time there was controversy as to whether a constructive proof was even possible~\cite{Hellman,Billinge,Richman,Richman2}. With hindsight,  disagreement on what methods are legitimately to be deemed ``constructive'' is the key point of the constructivist debate in the 1990s~\cite{Hellman,Billinge,Richman,Richman2}.
Even with modern constructive (in principle) proofs, the construction is not particularly explicit, and often very little is said as to what the quantum density matrix actually looks like. Traditionally the analysis stops, and the theorem is complete, once the existence of the quantum density matrix $\rho$ is established.

Herein we will have very little to say about the theorem and proof themselves, focussing more on the implications:  We shall say a little more about the density matrix itself --- and shall provide two constructions (one implicit, one explicit) for the quantum density matrix $\rho$ in terms of the probabilities assigned to rays in the Hilbert space.

\section{Statement of Gleason's theorem}

An explicit statement of Gleason's theorem runs thus~\cite{Gleason}:

{\bf Theorem:} \\
\emph{
Suppose $H$ is a separable Hilbert space, (either real or complex). \\[5pt]
A measure on $H$ is defined to be a function $v(\cdot)$ that assigns a nonnegative real number to each closed subspace of $H$ in such a way that:
 If   $\{A_{i}\}$ is any countable collection of mutually orthogonal subspaces of $H$, and the closed linear span of this collection is $B$, then $v(B)=\sum _{i}v(A_{i})$. Furthermore we normalize to  $v(H)=1$. \\[5pt]
Then if the Hilbert space $H$ has dimension at least three, (either real or complex), every measure $v(\cdot)$ can be written in the form $v(A)=\mathrm{tr}(\rho \, P_{A})$, where $\rho$ is a positive semidefinite trace class operator with $\tr(\rho)=1$, and 
$P_{A}$ is the orthogonal projection onto~$A$.\hfill$\Box$
}

(Physicists would almost immediately focus on complex Hilbert spaces; but some of the mathematical literature also works with real Hilbert spaces.)

The original theorem gives one very little idea of what the density matrix might look like, and it is this topic we shall address. Indeed, the original theorem spends many pages proving that the valuation $v(P)$ uniformly continuous;  while this is certainly an extremely useful result, most physicists, (and applied mathematicians for that matter), would simply assume continuity on physical grounds.


%
%
%
\section{Elementary observations}

\enlargethispage{20pt}
Our first observation is that since $\rho$ is Hermitian we can diagonalize it and define
\begin{equation}
\rho = \sum_i \lambda_i \; Q_i.
\end{equation}
Here the $Q_i$ are taken to be 1-dimensional subspaces, and the $\lambda_i$ are to be repeated with the appropriate multiplicity.
Per Gleason's theorem, 
\begin{equation}
v (Q_j) = \tr(\rho Q_j) = \tr\left( \left[\sum_i \lambda_i  Q_i\right]\; Q_j \right) =  \sum_i \lambda_i \;\tr\left( Q_i \; Q_j\right) =  \lambda_j.
\end{equation}
So actually
\begin{equation}
\rho = \sum_i v(Q_i) \; Q_i,
\end{equation}
which does not (yet) help unless you can somehow extract the $Q_i$ in terms of the underlying valuation function $v(\cdot)$. 

Furthermore note that for each 1-dimensional subspace $Q_i$ we can identify
\begin{equation}
Q_i \sim |\psi_i\rangle \; \langle \psi_i|
\end{equation}
where $|\psi_i\rangle$ is any arbitrary vector in the 1-dimensional subspace $Q_i$. Then
\begin{equation}
v(Q_i) =  \langle \psi_i| \rho  |\psi_i\rangle.
\end{equation}

Now let $P_i$ be any arbitrary collection of orthogonal 1-dimensional projection operators
\begin{equation}
v\left( \sum_i P_i\right) =  \sum_i  v(P_i) = 1.
\end{equation}
Using Gleason's theorem, we can calculate
\begin{equation}
v(P_j) = \tr(\rho P_j) = \tr\left( \left[\sum_i v(Q_i)  Q_i\right]\; P_j \right) =  \sum_i v(Q_i) \; \tr\left( Q_i \; P_j\right) =  \sum_i v(Q_i) \; S_{ij},
\end{equation}
with $S_{ij}= \tr\left( Q_i \; P_j\right) $ a bi-stochastic matrix. That is,  Gleason's theorem implies
\begin{equation}
v(P_j) = \sum_i v(Q_i) \;  S_{ij}; 
\qquad\qquad\hbox{with} \qquad\qquad 
S_{ij} = | \langle q_i | p_j \rangle |^2 = |U_{ij}|^2.
\end{equation}
So we see that the matrix $S_{ij}$ is actually unitary-stochastic; both unitary and unitary-stochastic matrices drop out automatically.

\def\ss{\hbox{\large $\displaystyle\$ $}}
Now pick some random basis $P_i$ and construct
\begin{equation}
\rho_P = \sum_i v(P_i) \; P_i.
\end{equation}
This is not $\rho$ itself, but it is what you get from $\rho$ by hitting it with $\ss_P$, the decoherence super-scattering operator with respect to the basis $P_i$~\cite{ana}. To see this note
\begin{equation}
\ss_P \;\rho =  \sum_i P_i \;\tr( P_i \,\rho) =    \sum_i P_i \; v( P_i) = \rho_P.
\end{equation}

Finally consider what happens if you \emph{average} over the $P_i$:
\begin{equation}
\left\langle \ss_P\right\rangle \;\rho =  \left\langle \sum_i P_i \;\tr( P_i \,\rho) \right\rangle 
=    \left\langle \sum_i P_i \,\; v( P_i) \right\rangle = \left\langle \rho_P \right\rangle.
\end{equation}
In $d$ dimensions for a uniform average over the $(P_i)_{ab}$ we have
\begin{equation}
\left\langle \sum_i (P_i)_{ab} \; (P_i)_{cd} \right\rangle = {\delta_{ac}\delta_{bd} + \delta_{ab} \delta_{cd}\over d+1}.
\end{equation}
This arises from symmetry plus the normalization condition $\langle I_{d\times d}\rangle = I_{d\times d}$. But then we can reconstruct
\begin{equation}
\rho = (d+1) \left\langle \rho_P \right\rangle - I_{d\times d}. 
\end{equation}
(Note this does have the correct trace, $\tr(\rho)=1$.)
So if you know all possible ways in which the density matrix decoheres $\rho \to \rho_P$, and uniformly average over all choices of decoherence basis, then one can reconstruct the full density matrix.   
While certainly an elegant result, this is by no means explicit.


\section{Implicit construction for the density matrix}

Let us now set up a reasonably explicit construction of the density matrix $\rho$ directly from the valuation function $v(P)$.

To construct $\rho$ proceed as follows: First for any 1-dimensional subspace note $Q\sim |n\rangle\; \langle n|$ where $n$ can be taken to be a unit vector in $S^{d-1}$. This defines a valuation $v(n)$ on $S^{d-1}$. 
Then find a ${n_1}$ such that $v(Q_{n_1})= \max_{n\in S^{d-1}} \{v(P_n)\} =  \max_{n\in S^{d-1}} \langle n|\rho|n\rangle$. 

Now consider the $S^{d-2}$ perpendicular to $n_1$: Proceed as follows --- find a $n_2$ such that $v(Q_{n_2})= \max_{n\in S^{d-2}} \{v(P_n)\}$. 
By construction $n_1 \perp n_2$ and $P_{n_1} P_{n_2} = 0$.

Iterate this construction: Consider the $S^{d-i}$ perpendicular to $n_1$, $n_2$, \dots, $n_{i-1}$: Find a $n_i$ such that $v(Q_{n_{i}})= \max_{n\in S^{d-i}} \{v(P_n)\}$. 
By construction the $n_j$ for $j\in\{1,2,\cdots, i\}$ are mutually perpendicular, and $P_{n_j} P_{n_k} = 0$ for $j\neq k$ and $j,k\in\{1,2,\cdots, i\}$.

Ultimately we have $n_d = \max_{n\in S^{0}} \{v(P_n)\} = \min_{n\in S^{d-1}} \{v(P_n)\}$.

\enlargethispage{25pt}
The construction terminates after $d$ steps with an orthonormal basis $n_1$, $n_2$, \dots, $n_d$,
and the corresponding valuations $v(Q_{n_i})$. Now construct
\begin{equation}
\rho = \sum_{i=1}^d v(Q_{n_i}) \; Q_{n_i}.
\end{equation}
This is the density matrix you want. \hfill $\Box$

{\bf Proof:}\\
It is clearly \emph{a} density matrix; it only remains to check that it is \emph{the} density matrix.

But this is obvious from the construction --- the $n_i$ are the simply eigenvectors of $\rho$, with the corresponding projection operators $Q_{n_i}$, and the $v(Q_{n_i})$ are the eigenvalues.  
(Basically the construction above is just an application of the  Rayleigh--Ritz min-max variational theorem for  finding eigenvectors/eigenvalues of Hermitian matrices.) The density matrix is constructed in terms of the values, $v(Q_{n_i})$, and locations, $n_i$, of the maximum, minimum, and extremal points of the valuation function $v(\cdot)$.  \hfill $\Box$

Note the construction is still rather implicit. Once Gleason's theorem guarantees the existence of the density matrix, this construction implicitly allows one to determine the density matrix. The more purist of constructivist mathematicians might not call this constructive, but most others would. 
On the other hand, as we shall now show, much better can be done in terms of a fully explicit construction.

\section{Explicit construction for the density matrix}

This second construction is completely explicit but considerably more subtle.
We assert that within the framework of Gleason's theorem, for any arbitrary basis on complex Hilbert space we can write:
\begin{eqnarray}
&&\rho =  \sum_j |n_j\rangle \; v(n_j) \; \langle n_j|
\\
&&+{1\over2} \sum_{j\neq k}  |n_j\rangle
\left\{ {v\left(n_j+n_k\over\sqrt{2}\right) - v\left(n_j-n_k\over\sqrt{2}\right)} -i \; {v\left(n_j+in_k\over\sqrt{2}\right)+iv\left(n_j-in_k\over\sqrt{2}\right)}\right\}     
 \; \langle n_k|.
 \nonumber
\end{eqnarray}
That is, to reconstruct the full density matrix we need only determine the valuations $v(\cdot)$, which is a collection of real numbers, on the specific set of unit vectors
\begin{equation}
n_j; \qquad \left(n_j\pm n_k\over\sqrt{2}\right); \qquad \left(n_j\pm in_k\over\sqrt{2}\right).
\end{equation}
There are a total of $d + d(d-1) + d(d-1) = 2d^2-d$ such unit vectors to deal with.

\enlargethispage{30pt}
This formula for the density matrix can also be rearranged as follows
\begin{eqnarray}
\rho &=&  \sum_j  v(n_j) |n_j\rangle \; \langle n_j| 
\nonumber\\
&&+{1\over2} \sum_{j< k}  
\left\{ {v\left(n_j+n_k\over\sqrt{2}\right) - v\left(n_j-n_k\over\sqrt{2}\right)} \right\}   
\left( \vphantom{\Big{|}} |n_j\rangle \; \langle n_k| +  |n_k\rangle \; \langle n_j| \right)  
\nonumber\\
&&-{i\over2} \sum_{j< k}  
\left\{  \; {v\left(n_j+in_k\over\sqrt{2}\right)+v\left(n_j-in_k\over\sqrt{2}\right)}\right\}    
\left( \vphantom{\Big{|}} |n_j\rangle \; \langle n_k| -  |n_k\rangle \; \langle n_j| \right).  
\end{eqnarray}
In this form, Hermiticity of the density matrix is manifest.

The situation for a real Hilbert space is considerably simpler:
\begin{equation}
\rho =  \sum_j |n_j\rangle \; v(n_j) \; \langle n_j|
+{1\over2} \sum_{j\neq k}  |n_j\rangle
\left\{ {v\left(n_j+n_k\over\sqrt{2}\right) - v\left(n_j-n_k\over\sqrt{2}\right)} \right\}     
 \; \langle n_k|.\qquad
\end{equation}
There are now only a total of $d + d(d-1) = d^2$ unit vectors to deal with.

This formula for the (real) density matrix can also be rearranged as follows
\begin{eqnarray}
\rho &=&  \sum_j  v(n_j) |n_j\rangle \; \langle n_j| 
\nonumber\\
&&+{1\over2} \sum_{j< k}  
\left\{ {v\left(n_j+n_k\over\sqrt{2}\right) - v\left(n_j-n_k\over\sqrt{2}\right)} \right\}   
\left( \vphantom{\Big{|}} |n_j\rangle \; \langle n_k| +  |n_k\rangle \; \langle n_j| \right).
\end{eqnarray}
In this form, symmetry of the (real) density matrix is manifest.

To start the construction, following Richman and Bridges~\cite{Richman},  we extend the valuation $v(P)\longleftrightarrow v(n)$ from $S^{d-1}$ to all of $H$ as follows:
\begin{equation}
f(n) =  ||n||^2 \; v\left( n\over ||n||\right),
\end{equation}
Now, again following Richman and Bridges, define~\cite[pages 2 and 7]{Richman},
\begin{equation}
\langle x| \rho| y \rangle = {f(x+y) - f(x-y)\over 4} -i \; {f(x+iy)-f(x-iy)\over 4},
\end{equation}
which in the real case reduces to
\begin{equation}
\langle x| \rho| y \rangle = {f(x+y) - f(x-y)\over 4}.
\end{equation}
Richman and Bridges~\cite[page 8]{Richman} assert the equivalent of:
\begin{itemize}
\itemsep0pt
\item $ \langle a x| \rho| b y \rangle  = \overline{a}\,b\, \langle x| \rho| y \rangle $.
\item $\langle x| \rho| y \rangle  = \overline{ \langle y| \rho| x \rangle }$.
\item $\langle x| \rho| y_1+y_2 \rangle = \langle x| \rho| y_1 \rangle + \langle x| \rho| y_2 \rangle$.
\end{itemize}
This is needed to verify that $\langle x| \rho| y \rangle $ actually represents a bilinear form.

Then the density matrix $\rho$ can itself be defined by
\begin{equation}
\rho =  \sum_j \sum_k |n_j\rangle \; \langle n_j| \rho| n_k \rangle \; \langle n_k|.
\end{equation}
So
\begin{equation}
\rho =  \sum_j \sum_k |n_j\rangle \; 
\left\{ {f(n_j+n_k) - f(n_j-n_k)\over 4} -i \; {f(n_j+in_k)-f(n_j-in_k)\over 4}\right\}     
 \; \langle n_k|.
\end{equation}
Whence, splitting the sum into diagonal and off-diagonal pieces, and noting that both $||n_j\pm n_k||^2 = 2 = ||n_j\pm in_k||^2$, while $ \widehat{n_j\pm n_k} = (n_j\pm n_k)/\sqrt 2$, and finally $ \widehat{n_j\pm i n_k} =  (n_j\pm i n_k)/\sqrt 2$, we have:
\begin{eqnarray}
&&\rho =  \sum_j |n_j\rangle \; v(n_j) \; \langle n_j|
\\
&&+{1\over2} \sum_{j\neq k}  |n_j\rangle
\left\{ {v\left(n_j+n_k\over\sqrt{2}\right) - v\left(n_j-n_k\over\sqrt{2}\right)} -i \; {v\left(n_j+in_k\over\sqrt{2}\right)+iv\left(n_j-in_k\over\sqrt{2}\right)}\right\}     
 \; \langle n_k|.
 \nonumber
\end{eqnarray}

That is, in terms of the decohered density matrix $\rho_P$ we have:
\begin{eqnarray}
&&
\rho =  \rho_P
\\
&&
+{1\over2} \sum_{j\neq k}  |n_j\rangle
\left\{ {v\left(n_j+n_k\over\sqrt{2}\right) - v\left(n_j-n_k\over\sqrt{2}\right)} -i \; {v\left(n_j+in_k\over\sqrt{2}\right)+iv\left(n_j-in_k\over\sqrt{2}\right)}\right\}     
 \; \langle n_k|.
 \nonumber
\end{eqnarray}
For a real Hilbert space this reduces to
\begin{equation}
\rho =  \rho_P
+{1\over2} \sum_{j\neq k}  |n_j\rangle
\left\{ {v\left(n_j+n_k\over\sqrt{2}\right) - v\left(n_j-n_k\over\sqrt{2}\right)} \right\}     
 \; \langle n_k|.
\end{equation}
One aspect of the ``miracle'' of Gleason's theorem is  that this construction is actually independent of the specific basis chosen.

To see why this construction works, note that from Gleason's theorem, for unit vectors 
\begin{equation}
\hat x \sim |\hat x\rangle ={|x\rangle\over ||x||}\sim {x\over||x||},
\end{equation}
we have
\begin{equation} 
v(\hat x) = \langle \hat x | \rho |\hat x\rangle = { \langle  x | \rho |x\rangle\over ||x||^2},
\end{equation} 
or more prosaically
\begin{equation} 
 \langle  x | \rho |x\rangle = ||x||^2 v(\hat x).
\end{equation} 
But then
\begin{equation} 
 \langle  x+y | \rho |x+y\rangle = ||x+y||^2 v(\widehat{x+y})  = 
  \langle x | \rho |x\rangle +  \langle  y | \rho |y\rangle 
   + ( \langle  x | \rho |y\rangle  + \langle  y | \rho |x\rangle ),
\end{equation} 
and 
\begin{equation} 
 \langle  x-y | \rho |x-y\rangle = ||x-y||^2 v(\widehat{x-y})  = 
  \langle x | \rho |x\rangle +  \langle  y | \rho |y\rangle 
   - ( \langle  x | \rho |y\rangle  + \langle  y | \rho |x\rangle ),
\end{equation} 
whence
\begin{equation} 
 \langle  x | \rho |y\rangle  + \langle  y | \rho |x\rangle
 = {1\over2} \left\{||x+y||^2 v(\widehat{x+y}) - ||x-y||^2 v(\widehat{x-y})  \right\}.
 \end{equation} 
(In a real Hilbert space we could stop here since then $\langle  x | \rho |y\rangle  = \langle  y | \rho |x\rangle$.)
 
 Similarly, in a complex Hilbert space, 
 \begin{equation} 
 \langle  x+iy | \rho |x+iy\rangle = ||x+iy||^2 v(\widehat{x+iy})  = 
  \langle x | \rho |x\rangle +  \langle  y | \rho |y\rangle 
   + i ( \langle  x | \rho |y\rangle  - \langle  y | \rho |x\rangle ),
\end{equation} 
and
\begin{equation} 
 \langle  x-iy | \rho |x-iy\rangle = ||x-iy||^2 v(\widehat{x-iy})  = 
  \langle x | \rho |x\rangle +  \langle  y | \rho |y\rangle 
   - i( \langle  x | \rho |y\rangle  - \langle  y | \rho |x\rangle ),
\end{equation} 
whence
\begin{equation} 
 \langle  x | \rho |y\rangle  - \langle  y | \rho |x\rangle
 = -{i\over2} \left\{||x+iy||^2 v(\widehat{x+iy}) - ||x-iy||^2 v(\widehat{x-iy})  \right\}.
 \end{equation} 
 Combining these results
 \begin{eqnarray} 
 \langle  x | \rho |y\rangle &=&
 +  {1\over4} \left\{||x+y||^2 v(\widehat{x+y}) - ||x-y||^2 v(\widehat{x-y})  \right\}
 \nonumber
  \\
 && -{i\over4} \left\{||x+iy||^2 v(\widehat{x+iy}) - ||x-iy||^2 v(\widehat{x-iy})  \right\}.
 \end{eqnarray} 
 This finally justifies our construction of the density matrix $\rho$ as presented above.

\section{Two dimensions}

Although Gleason's theorem does not  apply in two dimensions, there are improved versions of Gleason's theorem based on POVMs (positive operator valued measures), see~\cite{Busch,Caves}, that do apply to 2-dimensional Hilbert space. In this case the formalism simplifies even further: Let $\hat x$ and $\hat y$ be any orthonormal basis for the 2-dimensional Hilbert space. Then in terms of the valuation $v(\cdot)$ the density matrix is
\begin{eqnarray}
\rho &=&  v(\hat x) \; |\hat x\rangle \; \langle \hat x| +  v(\hat y) \; |\hat y\rangle \; \langle \hat y| 
\nonumber\\
&&+{1\over2} 
\left\{ {v\left(\hat x+\hat y\over\sqrt{2}\right) - v\left(\hat x-\hat y\over\sqrt{2}\right)} \right\}   
\left( \vphantom{\Big{|}} |\hat x\rangle \; \langle \hat y| +  |\hat y\rangle \; \langle \hat x| \right)  
\nonumber\\
&&-{i\over2}  
\left\{  \; {v\left(\hat x+i\hat y\over\sqrt{2}\right)-v\left(\hat x-i\hat y\over\sqrt{2}\right)}\right\}    
\left( \vphantom{\Big{|}} |\hat x\rangle \; \langle \hat y| -  |\hat y\rangle \; \langle \hat x| \right).  
\end{eqnarray}

If desired one can further rewrite this in terms of the Pauli $\sigma$ matrices
\begin{eqnarray}
\rho &=&  {v(\hat x) + v(\hat y)\over2} \; I_{2\times2} + {v(\hat x) - v(\hat y)\over2}\; \sigma_z 
\nonumber\\
&+&{1\over2} 
\left\{ {v\left(\hat x+\hat y\over\sqrt{2}\right) - v\left(\hat x-\hat y\over\sqrt{2}\right)} \right\}   \sigma_x
-{i\over2}  
\left\{  \; {v\left(\hat x+i\hat y\over\sqrt{2}\right)-v\left(\hat x-i\hat y\over\sqrt{2}\right)}\right\}    \sigma_y.
\qquad
\end{eqnarray}
For real 2-dimensional Hilbert space this further simplifies to
\begin{eqnarray}
\rho &=&  v(\hat x) \; |\hat x\rangle \; \langle \hat x| +  v(\hat y) \; |\hat y\rangle \; \langle \hat y| 
\nonumber\\
&&+{1\over2} 
\left\{ {v\left(\hat x+\hat y\over\sqrt{2}\right) - v\left(\hat x-\hat y\over\sqrt{2}\right)} \right\}   
\left( \vphantom{\Big{|}} |\hat x\rangle \; \langle \hat y| +  |\hat y\rangle \; \langle \hat x| \right).  
\end{eqnarray}
(For completeness, note that for one dimension the valuation trivializes to $v(\cdot)\equiv 1$, and so the density matrix trivializes to $\rho \equiv I_{1\times 1}$.)

\section{Discussion}

We have not attempted to provided a new proof of Gleason's theorem. Instead we have in mind a much more modest attempt at trying to understand what the density matrix actually looks like directly in terms of the probability valuations $v(\cdot)$ on a limited number of subspaces of the Hilbert space.

Gleason's theorem is profound that it shapes the probabilistic nature of quantum theory.  It places strong constraints on any attempts to modify this formalism, and it also gives a fundamental reason for why density operators play such an important role.  A vast amount of literature has been accrued on Gleason's theorem and its applications.  Many physicists and mathematicians have tried to simplify the proof and extend it to more generalized structures.  For a complete treatment, refer to the monograph by Hamhalter~\cite{Hamhalter}.

Future work regarding this explicit construction of the density operator may involve applications to quantum information theory.  This may reveal interesting links between quantum foundations, and to the fundamental results of quantum information theory such as no-cloning or no-broadcasting~\cite{BR}.  Such a direction would allow the reach of Gleason's theorem to extend further into the modern information-theoretic setting of quantum physics.

\acknowledgments{
DR is indirectly supported by the Marsden fund, \\
administered by the Royal Society of New~Zealand.
\\
MV is directly supported by the Marsden fund, \\
administered by the Royal Society of New~Zealand.}

\bigskip
\hrule
\vspace{-10pt}
\section*{Background resources}

For a general introduction to Gleason's theorem, see:
\begin{itemize}
\item[$\bullet$]
\url{https://en.wikipedia.org/wiki/Gleason's_theorem}

\item[$\bullet$]
\url{https://ncatlab.org/nlab/show/Gleason's+theorem}

\item[$\bullet$]
\url{https://plato.stanford.edu/entries/qt-quantlog/}

\end{itemize}

\bigskip
\hrule
\vspace{-10pt}
 
\end{document}